\documentclass[VANCOUVER,Times2COL]{WileyNJDv5} 

\usepackage{tabularray}
\usepackage{subcaption}
\usepackage{graphicx}

\articletype{ORIGINAL RESEARCH}%

{\color{black}
\received{Date Month Year}
\revised{Date Month Year}
\accepted{Date Month Year}
\journal{IET Renewable Power Generation}
\volume{00}
\copyyear{2023}
\startpage{1}
}

\begin{document}

\title{Federated Learning Assisted Distributed Energy Optimization}

\author[1]{Yuhan Du}

\author[2]{Nuno Mendes}

\author[2]{Simin Rasouli}

\author[1]{Javad Mohammadi}

\author[2]{Pedro Moura}

\authormark{DU \textsc{et al.}}
\titlemark{Federated Learning Assisted Distributed Energy Optimization}

\address[1]{\orgdiv{Department of Civil, Architectural and Environmental Engineering}, \orgname{University of Texas at Austin}, \orgaddress{\city{Austin}, \state{TX}}}

\address[2]{\orgdiv{Department of Electrical and Computer Engineering}, \orgname{University of Coimbra}, \orgaddress{\city{Coimbra}, \country{Portugal}}}

\corres{Javad Mohammadi, \orgdiv{Department of Civil, Architectural and Environmental Engineering}, \orgname{University of Texas at Austin}, \orgaddress{\city{Austin}, \state{TX}}. \email{javadm@utexas.edu}}



\abstract[Abstract]{

The increased penetration of distributed energy resources and the adoption of sensing and control technologies are driving the transition from our current centralized electric grid to a distributed system controlled by multiple entities (agents). The Transactive Energy Community (TEC) serves as an established example of this transition. Distributed energy management approaches can effectively address the scalability, resilience, and privacy requirements of the evolving grid.
In this context, the accuracy of agents' estimations becomes crucial for the performance of distributed and multi-agent decision-making paradigms. This paper specifically focuses on integrating Federated Learning (FL) with the multi-agent energy management procedure. FL is utilized to forecast agents' local energy generation and demand, aiming to accelerate the convergence of the distributed decision-making process.
To enhance energy aggregation in TECs, we propose an FL-assisted distributed \textit{Consensus + Innovations} approach. The results demonstrate that employing FL significantly reduces errors in predicting net power demand. The improved forecast accuracy, in turn, introduces less error in the distributed optimization process, thereby enhancing its convergence behavior.
}

\keywords{
\textit{Consensus + Innovations} approach, distributed optimization, federated learning, multi-agent systems, smart grid, transactive energy community
}


\maketitle



\section*{Nomenclature}\label{s:nomenclature}
\begin{table}[h!]
\begin{tabular}[h!]{lp{6.7cm}}


$g$& VPP index \\

$G$& Set of VPPs in the system\\

$N_G$& Number of VPPs\\

$C_{g}^{g2c}(P_{g})$& Electricity cost of exporting electricity from VPP $g$ to the community \\

$C_{g}^{c2g}(P_{g})$& Electricity cost of exporting electricity from the community to VPP $g$ \\

$c_{1, g}^{g2c}$, $c_{2, g}^{g2c}$& Cost function parameters for VPP $g$ to export electricity to the community \\

$c_{1, g}^{c2g}$, $c_{2, g}^{c2g}$& Cost function parameters for the community to export electricity to VPP $g$  \\

$P_{g}$& Power net demand of VPP $g$ \\

$P_{g}^*$& Power net demand of VPP $g$ at optimal solution \\

$P_{community}$& Net demand of the community \\

$\overline{P_{g, g2c}}$& The maximum possible power VPP $g$ can export to the community\\

$\overline{P_{g, c2g}}$& The maximum possible power from the community that VPP $g$ can consume\\

$\lambda^*$& System energy price at optimal solution\\

$\lambda_g^t$& Locational marginal price of VPP $g$ at iteration $t$\\

$P_{g}^t$& Power net generation of VPP $g$ at iteration $t$\\

$\mathcal{B}_r$& Set of buildings chosen by the server to participate in a FL round \\



\end{tabular}
\end{table}

\begin{table}[h!]
\begin{tabular}[h!]{lp{6.7cm}}
 \\
 \\
 \\


$\mathcal{K}$& Number of buildings allowed to participate in a FL round \\

$\mathcal{W}_a$& Weights computed by the FL server \\

$\mathcal{W}^b_{r}$& Weights that will be sent to the server by building $b$, in round $r$ \\

$\Omega_{\underline{B}}$& Set of VPPs reaching minimum net generation limits in the energy aggregation problem\\

$\Omega_{\overline{B}}$& Set of VPPs reaching maximum net generation limits in the energy aggregation problem \\

$\mu_{u, g}$& Inequality Lagrangian multiplier for upper power generation bound of VPP $g$\\

$\mu_{l, g}$& Inequality Lagrangian multiplier for lower power generation bound of VPP $g$\\

$i$& Agent index \\

$I$& Set of agents in the system\\

$N_I$& Number of agents\\



$\beta_i^{t+1}$, $\alpha_i^{t+1}$& Tuning parameters of agent $i$ at iteration $t+1$ \\

$\Omega_i$& Set of agent $i$'s neighbors \\

$\lambda_{i, \tau}^{conv}$& Converged locational energy price for agent $i$ for time period $\tau$\\

$\lambda_{\tau}^*$& System energy price at optimal solution for time period $\tau$\\

$N_m$& Maximum number of iterations for \textit{Consensus + Innovations} process \\

$f$& Objective function value \\

$f^*$& Optimal objective function value \\

\end{tabular}
\end{table}

\section{Introduction} \label{s:intro}

This section begins by outlining the motivation for integrating Federated Learning (FL) into energy forecasting, highlighting the significance of accurate forecasting within the distributed energy optimization framework, particularly the \textit{Consensus + Innovations} approach. 
Subsequently, it reviews relevant literature in the field and highlights the contributions of the present work. 
Finally, it describes the structure of the remainder of the paper.

\subsection{Motivation} \label{s:intro_motivation}
According to GridWise Architecture Council~\cite{TE_definition}, transactive energy is ``a system of economic and control mechanisms that allows the dynamic balance of supply and demand across the entire electrical infrastructure using value as a key operational parameter''. Therefore, the Transactive Energy Community (TEC) can be described as a group of connected buildings that use prices and incentives to guarantee economic benefits for all members. 
In the TEC context, these signals can be controlled to enhance the alignment between demand and generation, facilitating the development of a net-zero-energy community.

Management systems of TECs rely on the forecast of the energy profiles to maximize the local generation's self-consumption, thereby minimizing energy costs. For example, in scenarios where the generation is higher than the demand, it is possible to incentivize the TEC members to buy energy in that period by decreasing the tariffs in the community. Therefore, building management systems need to predict such price signals in advance to manage energy consumption and minimize energy bills.

Characterized by numerous decentralized and autonomous buildings, a TEC is suitable for the structure of a multi-user (agent) system.
Multi-user (agent) paradigms lend themselves well to managing and orchestrating a diverse set of autonomous buildings of a TEC.
In this setup, each user (agent) can execute local computations and engage in communication~\cite{du2022learning}. 
Multi-agent optimization techniques create a collaborative framework where agents collectively solve problems (e.g., maximizing self-consumption) by utilizing local computational resources and communication networks
\cite{kar2014distributed}.
In this paper, we utilize a \textit{Consensus + Innovations} approach~\cite{kar2012distributed} to regulate the collaboration between members of a TEC.
In this approach, the \textit{Consensus} term enforces an agreement between agents, while the \textit{Innovations} term guides local constraints fulfillment. 
Local constraints are associated with the local buildings' net demand, which requires an accurate forecast~\cite{kar2012distributed}, and FL can address this need.

FL's architecture allows a collaborative and privacy-preserving training of Machine Learning (ML) models. In the TEC context, FL allows the use of private data, like the number of users in the building, to achieve more accurate forecast models. Additionally, collaboration enables ML models to gain more adaptivity, making the accuracy levels more stable. For example, in new TECs, where the buildings can have insufficient samples, it is possible to take advantage of a pre-trained ML model by a TEC that has been using a FL framework. 

This study uses an FL-assisted \textit{Consensus + Innovations} approach to address the multi-agent distributed energy aggregation problems within a TEC. 
The FL framework enhances the forecast of net power demand for individual buildings of a TEC. 
A significant benefit of the FL approach is its capacity to augment forecast model training with private data, coupled with the capability to ``transfer'' the ML model to different communities, thereby preserving privacy and improving model robustness across varied datasets.
The benefits of FL prediction are more pronounced in the TECs with an increasing number of Virtual Power Plants (VPPs).
VPPs serve as collective entities of Distributed Energy Resources (DERs), forming a crucial link between TEC and the main electric grid~\cite{li2023machine}. 
VPPs can generate and consume electrical power; hence, predicting their net output is challenging.
From a multi-agent system perspective, VPPs are recognized as autonomous agents with communication and data processing capabilities. In this work, we aim to integrate FL with VPP's autonomy.

\subsection{Related Works} \label{s:intro_related_works}
Scholars have extensively explored various distributed optimization algorithms tailored for multi-agent distributed systems~\cite{kargarian2016toward}. 
Li~\cite{li2023learning} introduced an approach to resolve the decentralized DC Optimal Power Flow (DCOPF) utilizing the Alternating Direction Method of Multipliers (ADMM), complemented by a data-driven strategy to expedite convergence. 
Muhammad~\cite{khan2022multi} innovated an agent-based distributed Energy Management system (ADEMS) at the microgrid scale to build a multi-agents-based optimal Electric Vehicle (EV) charging mechanism. 
Others~\cite{nguyen2021distributed}~\cite{jasim2023consensus} have developed consensus-based methodologies to evaluate agent performance in cyber-physical systems, also addressing network packet loss issues within multi-agent microgrids. 
Kar \cite{kar2012distributed} developed a novel version of consensus-based distributed algorithms called \textit{Consensus + Innovations} approach, wherein the \textit{Consensus} term ensures inter-agent agreements, while the \textit{Innovation} terms aim to satisfy local constraints.
This framework has been widely used for solving various optimization problems, such as Economic Dispatch (ED)~\cite{kar2014distributed}, DCOPF~\cite{mohammadi2014distributed}, Security Constrained Optimal Power Flow~\cite{mohammadi2016agent}, and EV charging issues~\cite{mohammadi2016fully,mohammadi2023towards}. 

In coordinated energy dispatch problems (such as TEC management), agents' forecasts play a key role in the distributed optimization procedure~\cite{ahmadifar_development_2023, dinh_milp-based_2022}. 
%
According to Kim and Ghalehkhondabi~\cite{kim_convolutional_2017, ghalehkhondabi_overview_2017}, artificial neural networks can provide a robust approach to approximating real-valued target functions and are the main ML method used for forecasting demand and generation in buildings. A framework was developed by Chae~\cite{chae_artificial_2016} to forecast with a day-ahead time horizon, in a 15-minute time sample. Moreover, Azadeh ~\cite{azadeh_annual_2006} developed a model for long-term forecasting where the target is the annual electricity consumption in high-energy buildings. Beyond the need for the forecast of electricity demand, the TEC management system also needs to know the forecast of local generation for prosumers. To this end, Sabzehgar and Zhang ~\cite{sabzehgar_solar_2020, zhang_forecast_2018} forecast the solar output.

Several studies have explored the use of FL to predict building energy profiles in multi-agent networks while protecting privacy and preserving users' data locally. FL can be utilized in a TEC~\cite{9960350}, to use local models to forecast energy demand, bringing beneficial advantages, such as allowing collaboration between agents without compromising data privacy. Fekri et al.~\cite{FEKRI2022107669} explored the challenges of traditional ML approaches that require data sharing with the server for training and suggested using FL as an alternative. The authors evaluated two FL strategies, FedSGD and FedAVG, and showed that FedAVG achieves better accuracy and requires fewer communication rounds. As discussed in~\cite{cheng2022review}, different types of FL methods can be applied to energy-related problems. Horizontal FL, vertical FL, and federated transfer learning are some of the classification criteria based on data partitioning in sample space and feature space. The structure and topology of the network also affect the classification of FL, leading to central server-based FL and distributed FL. In~\cite{9148937} FL was used for households' short-term load forecasting (STLF) of electricity.

Recent works have focused on addressing FL's shortcomings in energy consumption prediction, especially in predicting peak demand.
Clustering techniques have been proposed to improve the performance of FL in this area. Nightingale et al.~\cite{nightingale2022effect}, demonstrated that clustering data based on building types and weight updates can enhance the performance of FL. However, clustering can also create privacy risks depending on the method used. In~\cite{briggs2022federated} FL was compared with clustered variants, centralized and localized learning approaches, and showed that FL with a hierarchical clustering algorithm can improve performance and reduce computation time compared to localized learning. Another study~\cite{iot3030021} investigated the impact of different privacy levels of data on forecasting residential energy consumption scenarios using either central or distributed collecting data. They found that the accuracy of FL forecast is reduced as privacy levels increase, and a higher number of exchanging data in the network can pose a challenge for FL approaches. Gao et al.~\cite{10.1145/3485730.3493450} proposed PriResi, a privacy-preserved, communication-efficient, and cloud-service-free load forecasting system that relies on a decentralized FL framework. The authors showed that their approach achieves high prediction accuracy while preserving residents' privacy. Similarly, Savi and Olivadese~\cite{9469923} presented an architecture that combines FL and Edge Computing to improve the accuracy of short-term energy consumption forecasting while keeping sensitive data local and preserving users' privacy.

\subsection{Contribution} \label{s:intro_contribution}
This paper establishes a novel Federated Learning-assisted \textit{Consensus + Innovations} approach to optimize energy transactions between a TEC and VPPs. In this multi-pronged approach, the FL model is used to forecast the net energy demand for buildings of a TEC. Locally, the buildings forecast model will predict, demand and generation (if exist) independently. After this step, the forecasted net energy demand (the difference between local demand and generation) is shared with the community. This forecast will fed into the \textit{Consensus + Innovations} approach to calculate more reliable dispatch decisions (including system energy price and agent's power output).

This paper's contributions can be summarized as follows:

\begin{enumerate}
    \item Introducing a novel FL-assisted \textit{Consensus + Innovations} approach.
    \item Implementing a market system that will maximize TEC's self-consumption, by allowing transactions with VPPs.
    \item Forecasting the net power demand of a community by using an FL framework that allows reduced energy prediction error in complex scenarios, like limited record of data samples in buildings.
    \item Mitigating compounding error by reducing forecast errors leads that can adversely impact the convergence of distributed optimization process.
\end{enumerate}

\subsection{Paper Organization} \label{s:intro_paper_organization}
The rest of the paper is organized as follows: 
the mathematical models of the Energy Aggregation problem, the FL model for predicting net power demand, and the \textit{Consensus + Innovations} multi-agent solution approach are presented in Section II. 
Section III describes the details of the simulation case studies, including the data source, and setup of the FL and \textit{Consensus + Innovations} methods.

Section IV presents the forecasting results from FL and the simulation results of the \textit{Consensus + Innovations} approach. 
Section V brings the paper to a close, and the Appendix presents the derivation of analytical solutions for Energy Aggregation problems, as well as the derivation of FL in building energy profile predictions.

\section{Mathematical Models}  \label{s:model}

This section introduces the mathematical models of the Energy Aggregation problem, the FL model for predicting energy profiles in multi-agent systems, and the \textit{Consensus + Innovations} multi-agent solution approach. Lastly, it also introduces how both methodologies are seamlessly integrated.

\subsection{Energy Aggregation Problem} \label{s:model_ea}
\begin{figure}[htbp]
\centering
\includegraphics[width=0.9\columnwidth]{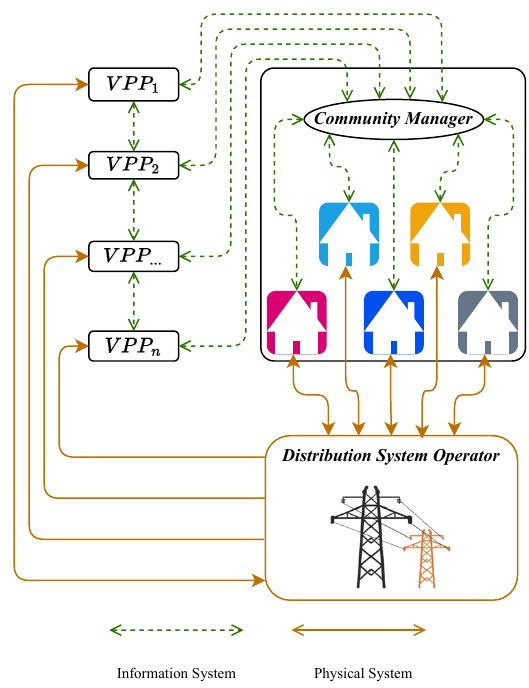}
\caption{Diagram of proposed distributed multi-agent power system infrastructure.}
\label{f:eap_diagram}
\end{figure}

A multi-agent distributed power system infrastructure that comprises a community and VPPs was considered, as illustrated in Fig. \ref{f:eap_diagram}.
Inside the community, all buildings have individual electric demand, and some buildings also have local generation.
The community manager manages the net demand within the community.
The community manager and VPPs can share information and local data processing. 
From the perspective of a multi-agent infrastructure, they are represented as agents.
The intercommunication between agents is facilitated through the established interconnections within the information system. Figure \ref{f:eap_diagram} showcases these entities' information and physical links.

In our multi-agent setup, the aggregation problem seeks to minimize the system-wide energy cost across the community and VPPs. 
The results of energy aggregation should respect the power demand-supply balance and fulfill the physical limitations of all entities.
Depending on the net demand of the community, the formulation varies between \eqref{eq:g2c1}-\eqref{eq:g2c3} (positive community net demand) and \eqref{eq:c2g1}-\eqref{eq:c2g3} (negative community net demand). 

The energy cost function parameters are set by VPPs, which vary over the time horizon and the net demand of the community. 
There are three conditions based on the net demand of the community.
(i) when the community has a positive net demand, the community imports power from VPPs and reimburses the VPPs.
Thus, the cost function coefficients are recognized as the ``selling price''.
In such conditions, \eqref{eq:g2c1}-\eqref{eq:g2c3} are used, and $P_g$ represents power exported from VPP $g$ to the community.
Note that some community buildings are equipped with photovoltaic (PV) and they may export the generation surplus to the community.
(ii) when the community has a negative net demand, the community exports power from VPPs and profits from this exchange.
The cost coefficients are considered as the ``buying price''.
In such conditions, \eqref{eq:c2g1}-\eqref{eq:c2g3} are used, and $P_g$ indicates power exported to the VPP $g$ from the community. 
(iii) when the community net demand is zero, there is no power flow between VPPs and the community.

The objective function in \eqref{eq:g2c1} minimizes the quadratic cost of exporting power from VPPs to the community. 
Likewise, \eqref{eq:c2g1} minimizes the quadratic cost of consuming power from the community to VPPs. 
The power balance within this multi-agent power system is achieved by \eqref{eq:g2c2} and \eqref{eq:c2g2}. 
Physical limits on the maximum available power that each VPP can export or import from the community are enforced by \eqref{eq:g2c3} and \eqref{eq:c2g3}.

\vspace{-0.4cm}
\begin{align}
    \min_{P_{g}} \quad & \sum_{g=1}^{N_{G}}C_{g}^{g2c}(P_{g})=\sum_{g=1}^{N_{G}}(c_{1, g}^{g2c} P_{g}^2+c_{2, g}^{g2c} P_{g}) \label{eq:g2c1} \\
     \textrm{s.t.} \quad & \sum_{g=1}^{N_{G}}P_{g}=P_{community} \label{eq:g2c2} \\
                         & 0 \leq P_{g}\leq \overline{P_{g}^{g2c}}, \forall g\in G \label{eq:g2c3}
\end{align}

\vspace{-0.4cm}
\begin{align}
    \min_{P_{g}} \quad & \sum_{g=1}^{N_{G}}C_{g}^{c2g}(P_{g})=\sum_{g=1}^{N_{G}}(c_{1, g}^{c2g} P_{g}^2+c_{2, g}^{c2g} P_{g}) \label{eq:c2g1} \\
     \textrm{s.t.} \quad & \sum_{g=1}^{N_{G}}P_{g}=-P_{community} \label{eq:c2g2} \\
                         & 0 \leq P_{g}\leq \overline{P_{g}^{c2g}}, \forall g\in G \label{eq:c2g3}
\end{align}

The Lagrange multipliers of \eqref{eq:g2c2} and \eqref{eq:c2g2} are the locational marginal prices ($\lambda_g$) for each VPP. 
After solving the energy aggregation problems, VPPs' $\lambda_g$ become identical and are referred to as the system energy price at the optimal solution ($\lambda^*$).
The process of finding a centralized analytical solution for this model is described in Appendix \ref{s:appendix_ea}. 
Under condition (i) (the community has a positive net demand), the optimal system energy price $\lambda^*$ and optimal power net generation $P_g^*$ for VPP $g$ are given by \eqref{eq:lamStar_g2c}, \eqref{eq:pg_non_g2c}, \eqref{eq:pg_upper_g2c} and \eqref{eq:pg_lower}. 
Likewise, under condition (ii) (the community has a negative net demand), the optimal system energy price $\lambda^*$ and optimal power net generation $P_g^*$ for VPP $g$ are given by \eqref{eq:lamStar_c2g} \eqref{eq:pg_non_c2g}, \eqref{eq:pg_upper_c2g} and \eqref{eq:pg_lower}. 



\vspace{-0.4cm}
\begin{multline} \label{eq:lamStar_g2c}
    \lambda^* = \left(\sum_{g\notin {\Omega_{\overline{B}}\cup\Omega_{\underline{B}}}}\frac{1}{2c_{1, g}^{g2c}}\right)^{-1} \left[P_{community} - \sum_{g\in {\Omega_{\overline{B}}}} \overline P_{g} \right. \\
    \left. \quad - \sum_{g\in {\Omega_{\underline{B}}}} \underline P_{g} + \sum_{g\notin{\Omega_{\overline{B}}\cup\Omega_{\underline{B}}}}\frac{c_{2, g}^{g2c}}{2c_{1, g}^{g2c}}\right]
\end{multline}

\vspace{-0.4cm}
\begin{multline} \label{eq:lamStar_c2g}
    \lambda^* = \left(\sum_{g\notin {\Omega_{\overline{B}}\cup\Omega_{\underline{B}}}}\frac{1}{2c_{1, g}^{c2g}}\right)^{-1} \left[-P_{community} - \sum_{g\in {\Omega_{\overline{B}}}} \overline P_{g} \right. \\
    \left. \quad - \sum_{g\in {\Omega_{\underline{B}}}} \underline P_{g} + \sum_{g\notin{\Omega_{\overline{B}}\cup\Omega_{\underline{B}}}}\frac{c_{2, g}^{c2g}}{2c_{1, g}^{c2g}}\right]
\end{multline}

\vspace{-0.4cm}
\begin{align}
    P_{g}^* = \frac{\lambda^* - c_{2, g}^{g2c}}{2c_{1, g}^{g2c}}, g\notin {\Omega_{\overline{B}}\cup\Omega_{\underline{B}}} \label{eq:pg_non_g2c} \\
    P_{g}^* = \frac{\lambda^* - c_{2, g}^{c2g}}{2c_{1, g}^{c2g}}, g\notin {\Omega_{\overline{B}}\cup\Omega_{\underline{B}}} \label{eq:pg_non_c2g} \\
    P_{g}^* = \overline{P_{g}^{g2c}}, g\in {\Omega_{\overline{B}}} \label{eq:pg_upper_g2c} \\
    P_{g}^* = \overline{P_{g}^{c2g}}, g\in {\Omega_{\overline{B}}} \label{eq:pg_upper_c2g} \\
    P_{g}^* = 0, g\in {\Omega_{\underline{B}}} \label{eq:pg_lower}
\end{align}

Here, ${\Omega_{\overline{B}}\cup\Omega_{\underline{B}}}$ refers to the set of VPPs not reaching the net generation limits.
Thus, $\lambda^*$ and $P_{g}^*$ can be analytically found for a centralized energy aggregation problem by \eqref{eq:lamStar_g2c}-\eqref{eq:pg_lower}. The model uses the optimal results from the centralized calculation as a baseline.

\subsection{Federated Learning Model} \label{s:model_fl}

In a TEC, forecast accuracy of the generation, demand, and net demand in each building is crucial to enhance energy management systems and, consequently, ensure objectives such as the minimization of energy costs or maximization of self-consumption. FL brings a new methodology that allows the use of personal and private data without the need to share it with other members. Therefore, using an FL approach, a TEC member can train their forecast model, and send the weights $\mathcal{W}^b_{r}$ relatively to the most promising local train iteration (epoch) to the TEC server. The TEC server, among a certain number of rounds $r = \{1, 2, \cdots, \mathcal{R}\}$, will receive by the $\mathcal{K}$ selected participants, $\mathcal{B}_r$, the respective computed weights. Thus, round by round, the TEC server will compute an average of the received weights, $\mathcal{W}_a$, using~\eqref{eq:FedAvg}. To start the new round, the TEC server will send the computed weights to the new participants $\mathcal{B}_{r+1}$. This process enables collaboration between all TEC members while protecting their private data.

\begin{equation}
    \label{eq:FedAvg}
    \mathcal{W}_a \gets \frac{1}{\mathcal{K}} \sum_{b=1}^{\mathcal{B}_r} \mathcal{W}^b_{r}
\end{equation}

\begin{figure}[htbp]
\centering
\includegraphics[width=1\columnwidth]{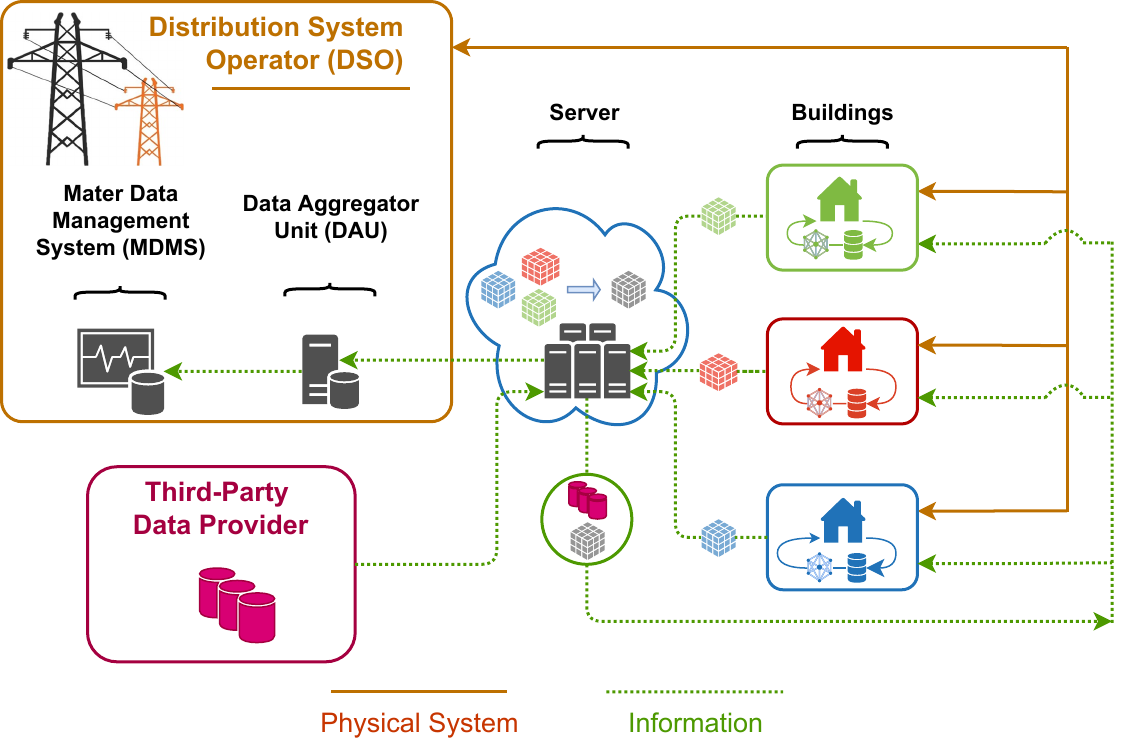}
\caption{Diagram of the developed FL framework for a transactive energy community.}
\label{f:fl_implementation}
\end{figure}

In this paper, we will leverage our prior work \cite{9960350} on FL algorithms tailored for TECs.
Beyond enhancing the forecast models by using their private data, the developed algorithm also adds a third party connected to the TEC server. This third party will supply weather data. From the perspective of the TEC members, such a strategy enhances their digital security by preventing their connections to external entities. The diagram of the implemented framework is presented in Figure~\ref{f:fl_implementation}. Green dotted lines represent the information flow, and orange lines represent the physical system of the power grid. 

Each TEC building, $b$, has two independent systems: demand forecaster and a forecast system for local PV generation. Both systems use the same ML architecture, represented in Table~\ref{t:lstm_arch}, which uses two Long-Short Term Models (LSTM), with the ReLu as an activation function, and two dropout layers.

\begin{table}[htpb]
\caption{Proposed LSTM architecture.}
\label{t:lstm_arch}
\centering
\begin{tabular*}{0.48\textwidth}{@{\extracolsep\fill}llll@{\extracolsep\fill}}
\toprule
\textbf{Layer} & \textbf{Type} & \textbf{LSTM cells} & \textbf{Activation Function}  \\ 
\midrule
LSTM           & Input         & 16                  & Relu                          \\
LSTM           & Hidden        & 16                  & Relu                          \\
Dropout(0.2)   & Hidden        & -                   & -                             \\
LSTM           & Hidden        & 16                  & Relu                          \\
Dropout(0.2)   & Hidden        & -                   & -                             \\
Dense          & Output        & -                   & -                             \\
\bottomrule
\end{tabular*}
\end{table}


\subsection{Consensus + Innovations Approach} \label{s:model_cpi}

The \textit{Consensus + Innovations} approach aims to solve the optimality conditions of the underlying optimization problem (energy aggregation) iteratively and in fully distributed fashion\cite{kar2014distributed}. 
Through local computations and communication with neighboring agents, agents cooperate to solve the energy aggregation problem \eqref{eq:g2c1}-\eqref{eq:g2c3} and \eqref{eq:c2g1}-\eqref{eq:c2g3}.  
At the beginning of each iteration, each agent distributes the local copy of the ``consensus variable" (i.e., local energy price $\lambda$) to neighboring agents. 
Then, the \textit{Consensus} term guides agreements among all agents while the \textit{Innovations} term guarantees that local conditions are fulfilled. 
To summarize, agents collaborate to reach a consensus on local copies of $\lambda$ while receiving the optimal nodal presumption value to conserve the supply and demand balance locally. 

\vspace{-0.4cm}
\begin{align}
    \lambda_i^{t+1} = \lambda_i^t - \beta_i^{t+1} \sum_{j\in\Omega_i}(\lambda_i^t - \lambda_j^t) - \alpha_i^{t+1}(P_{g}^t), i\in G  \label{eq:lam_vpp_1} \\
    \lambda_i^{t+1} = \lambda_i^t - \beta_i^{t+1} \sum_{j\in\Omega_i}(\lambda_i^t - \lambda_j^t) - \alpha_i^{t+1}(P_{community}), i\notin G  \label{eq:lam_cty_1} \\
    \lambda_i^{t+1} = \lambda_i^t - \beta_i^{t+1} \sum_{j\in\Omega_i}(\lambda_i^t - \lambda_j^t) - \alpha_i^{t+1}(-P_{g}^t), i\in G  \label{eq:lam_vpp_2} \\
    \lambda_i^{t+1} = \lambda_i^t - \beta_i^{t+1} \sum_{j\in\Omega_i}(\lambda_i^t - \lambda_j^t) - \alpha_i^{t+1}(P_{community}), i\notin G  \label{eq:lam_cty_2} \\
P_{g}^{t+1}=
\begin{cases}
    \frac{\lambda_g^{t+1} - c_{2, g}^{g2c}}{2c_{1, g}^{g2c}},& 0\leq \frac{\lambda_i^{t+1} - c_{2, g}^{g2c}}{2c_{1, g}^{g2c}}\leq \overline{P}_{G_i}\\
    \overline{P}_{G_i},                   & \frac{\lambda_i^{t+1} - c_{2, g}^{g2c}}{2c_{1, g}^{g2c}} > \overline{P}_{G_i}\\
    0,                               &\frac{\lambda_i^{t+1} - c_{2, g}^{g2c}}{2c_{1, g}^{g2c}} < 0
\end{cases} \label{eq:pg_1}\\
P_{g}^{t+1}=
\begin{cases}
    \frac{\lambda_g^{t+1} - c_{2, g}^{c2g}}{2c_{1, g}^{c2g}},& 0\leq \frac{\lambda_i^{t+1} - c_{2, g}^{c2g}}{2c_{1, g}^{c2g}}\leq \overline{P}_{G_i}\\
    \overline{P}_{G_i},                   & \frac{\lambda_i^{t+1} - c_{2, g}^{c2g}}{2c_{1, g}^{c2g}} > \overline{P}_{G_i}\\
    0,                               &\frac{\lambda_i^{t+1} - c_{2, g}^{c2g}}{2c_{1, g}^{c2g}} < 0
\end{cases} \label{eq:pg_2}
\end{align} 

The local optimization variables (local energy price $\lambda_g^t$ and VPP net generation $P_g^t$) are updated iteratively. 
Build on the foundation of our earlier work \cite{du2023need}, under condition (i) (positive net demand), the updates for $\lambda_g^t$ follow \eqref{eq:lam_vpp_1} and \eqref{eq:lam_cty_1}, and \eqref{eq:pg_1} determines $P_g^t$. 
Under condition (ii) (negative net demand), the updates for $\lambda_g^t$ follow \eqref{eq:lam_vpp_2} and \eqref{eq:lam_cty_2}, and \eqref{eq:pg_2} solves $P_g^t$. 
Community net demand appears in \eqref{eq:lam_cty_1} and \eqref{eq:lam_cty_2}, which is an estimation. 
A good estimation of community net demand boosts the performance of \textit{Consensus + Innovations} approach, which will be discussed later in Section \ref{s:results_cpi}.


Our earlier research \cite{kar2014distributed,mohammadi2016fully} demonstrate that \eqref{eq:lam_vpp_1}-\eqref{eq:pg_2} updates lead to the optimal solution when the communication topology is a connected graph and $\alpha_i^t$ and $\beta_i^t$ are appropriately tuned.


The pseudo-code for the \textit{Consensus + Innovations} approach is illustrated in Algorithm \ref{a:C+I}. 
Iteratively, agent $i$ executes the updates locally and exchanges the updated values of $\lambda_i$ with neighboring agents until the consensus on $\lambda$ is achieved and the optimality conditions of \eqref{eq:g2c1}-\eqref{eq:g2c3} and \eqref{eq:c2g1}-\eqref{eq:c2g3} are satisfied.
Without comparing with the optimal value, the convergence criterion was defined as the difference between the local $\lambda_i$ and $\lambda_j$ from neighboring agents $j\in \Omega_i$. As shown in \eqref{eq:conv}, this value is compared against a predefined threshold ($\varepsilon$). However, this criterion only ensures that agents reach a consensus, but this agreed value might be deviated. The results are presented and discussed in section \ref{s:results_cpi}.

\vspace{-0.4cm}
\begin{align}
\max|\lambda_i-\lambda_j| \leq \varepsilon, j\in \Omega_i, \forall i\in I  \label{eq:conv}
\end{align}

\vspace{-0.4cm}
\begin{algorithm}
\caption{\textit{Consensus + Innovations} Approach} 
\label{a:C+I}
\begin{algorithmic}[1]
\State \textbf{Agent-level initialization}: initialize $P_{i}$, $\lambda_i$, $\alpha_i$, and $\beta_i$
\For{$t=1, 2, ...$}
    \State Exchange $\lambda_i^t$ with neighboring agents
    \State Update $\lambda_i^{t+1}$ and $P_{G_i}^{t+1}$ by solving \eqref{eq:lam_vpp_1}-\eqref{eq:pg_2}
	\State \textbf{Terminate} if \eqref{eq:conv} holds
\EndFor
\end{algorithmic} 
\end{algorithm}

\subsection{Combined Framework} \label{s:model_flcpi}
\begin{figure}[htbp]
\centering
\setlength{\abovecaptionskip}{0.cm}
\includegraphics[width=1\columnwidth]{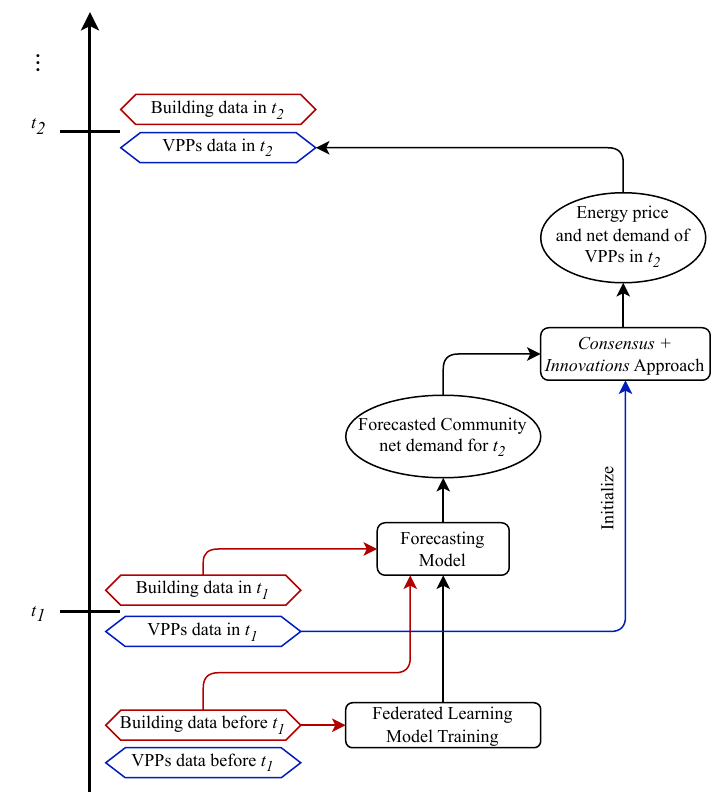}
\caption{Process of the FL-assisted \textit{Consensus + Innovations} approach framework when finding the energy price and net demand of VPPs in $t_2$.}
\label{f:simulation_flow}
\end{figure}

The real-time process for the FL-assisted \textit{Consensus + Innovations} approach is outlined in Figure \ref{f:simulation_flow}. 
This integrated method computes the energy pricing and net power demand of VPPs for $t_2$. 
Before $t_1$, an array of historical building data containing net power demand, weather, and geospatial data are utilized to train the FL model. 
Between $t_1$ and $t_2$, the data from these buildings is forwarded to the FL forecasting model, to forecast the community net demand for $t_2$. 
Concurrently, data relevant to the VPPs, such as cost function coefficients, power generation constraints, and energy price of $t_1$, are employed to initialize the \textit{Consensus + Innovations} algorithm. 
After integration of the forecasted net demand for the community, this algorithm resolves the fully distributed energy aggregation optimization problem, thereby deriving the optimal energy price and the net output of VPPs for $t_2$.

\section{Simulations}  \label{s:simulations}

This section details the simulation process for the FL-enhanced \textit{Consensus + Innovations} approach. 
It includes the sources of data utilized to train the FL model alongside the local models, as well as the configuration of the FL model. 
Then, this section illustrated the setup of the simulation, which encompasses the communication network and the procedural flow of the simulation.

\subsection{Data and Scenarios} \label{s:simulations_data} 


The National Renewable Energy Laboratory (NREL)'s public dataset of End-Use Load Profiles (EULPs) was used. This synthetic dataset includes simulated EULPs representing 550,000 residential and 350,000 commercial buildings across the United States.
To simulate a community of buildings, 125 residential buildings were selected. Such buildings were randomly selected from the San Diego region. All the selected buildings are prosumers, hence, requiring a forecast of the demand and the local generation to calculate the local net demand.

\begin{figure}[htpb]
    \centering
    \includegraphics[width=1\columnwidth]{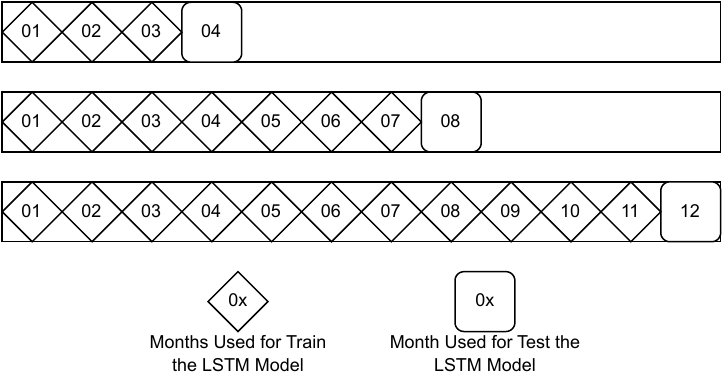}
    \caption{Diagram explanation of how the data was split for train and test.}
    \label{fig:data_split}
\end{figure}

To proceed to the tests using the FL framework, three different tests were implemented to embrace different seasonality. To maintain the time series sequence, it was decided that the achieved LSTM models for each test model needed to be trained with all previously available months. Then, the three months selected for the test were: (i) April, (ii) August, and (iii) December, as presented in Figure~\ref{fig:data_split}.

In order to obtain a fair comparison, the FL results were tested against a traditional methodology. This baseline simulates a community without FL, where the forecast ML models are only trained using local data. Additionally, the 125 selected buildings were split to form 2 communities: community $a$ and community $b$. Community $a$ is the primary and has 100 buildings. Community $b$, with the primary 25 buildings, aims to simulate the scenario of new buildings joining the main community. Therefore, for the buildings of the community $b$, it is considered that they have only data in the selected months for the test (April, August, and December), i.e., in this scenario, the buildings will train within the first 28 days and will forecast the last days of the month. The electricity net demand forecast of the community buildings $b$ will be performed with and without FL. In the test where FL will be used, in community $b$, a strategy that allows to start the training by using the last computed weights of community $a$ is implemented. 


\subsection{Federated Learning Setup \& Initialization}  \label{s:simulations_fl}

The setup for the initialization of the selected FL framework includes their parameters and the LSTM local hyperparameters. 
The implemented scenario is constituted of community $a$ (100 buildings) and community $b$ (25 buildings). 
All the $125$ buildings have a PV system installed. The selected FL framework allows training the forecast models independent from the selected system, i.e., in an FL round $\mathcal{R}$ a building with PV can be selected for training its forecast system of demand, but this not imply that its generation forecast model is also selected for training. Therefore, two independent parameterizations are needed, one for each system. For the global parameters of the FL architecture, behind the number of rounds, the number of participants in each round $\mathcal{K}$ needs to be defined. For the LSTM models, seven hyperparameters are needed: (i) local epochs $e$; (ii) batch size $b_s$; (iii) validation split $v_s$; (iv) past observation $p_o$; (v) future observations to be predicted $f_o$; (vi) number of cells in input layer $c_{il}$; and (vii) the number of cells in the hidden layers $c_{hl}$.  Table~\ref{t:fl_setup} presents the used parameters. It was defined that both systems will have the same configuration, only differing when it is community $a$ or community $b$. Additionally, to ensure fair comparison LSTM parameters are kept the same for the FL and the baseline model.

\begin{table}
\centering
\caption{Federated Learning parameters, and LSTM hyperparameters.}
\label{t:fl_setup}
\begin{tblr}{
  row{3} = {c},
  row{4} = {c},
  row{6} = {c},
  row{7} = {c},
  row{8} = {c},
  row{9} = {c},
  row{10} = {c},
  row{11} = {c},
  cell{1}{2} = {r=2}{c},
  cell{1}{3} = {c=2}{c},
  cell{2}{3} = {c},
  cell{2}{4} = {c},
  cell{3}{1} = {r=2}{},
  cell{5}{1} = {r=7}{},
  cell{5}{2} = {c},
  cell{5}{3} = {c=2}{c},
  cell{6}{3} = {c=2}{},
  cell{7}{3} = {c=2}{},
  cell{8}{3} = {c=2}{},
  cell{9}{3} = {c=2}{},
  cell{10}{3} = {c=2}{},
  cell{11}{3} = {c=2}{},
  hline{1,3,12} = {-}{0.08em},
  hline{5} = {-}{},
}
              & \textbf{Parameters} & \textbf{Community} &            \\
              &                     & \textit{a}         & \textit{b} \\
\textbf{FL}   & $\mathcal{R}$       & 30                 & 1          \\
              & $\mathcal{K}$       & 5                  & 25         \\
\textbf{LSTM} & $e$                 & 8                  &            \\
              & $b_s$               & 4                  &            \\
              & $v_s$               & 0.25               &            \\
              & $p_o$               & 96                 &            \\
              & $f_o$               & 1                  &            \\
              & $c_il$              & 16                 &            \\
              & $c_hl$              & 16                 &            
\end{tblr}
\end{table}

\begin{figure}[htpb]
    \centering
    \setlength{\abovecaptionskip}{0.cm}
    \includegraphics[width=0.5\linewidth]{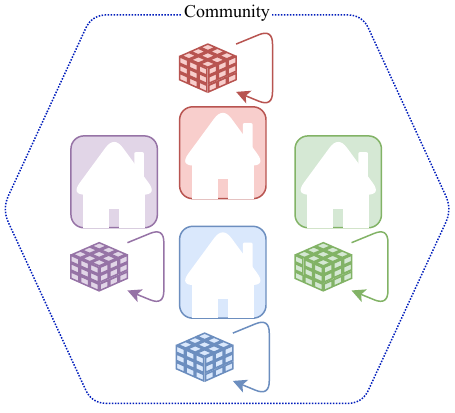}
    \caption{Diagram representing the local model training process for community $a$ and $b$.}
    \label{f:local_train_process}
\end{figure}

\begin{figure}[htpb]
    \centering
    \setlength{\abovecaptionskip}{0.cm}
    \includegraphics[width=0.8\linewidth]{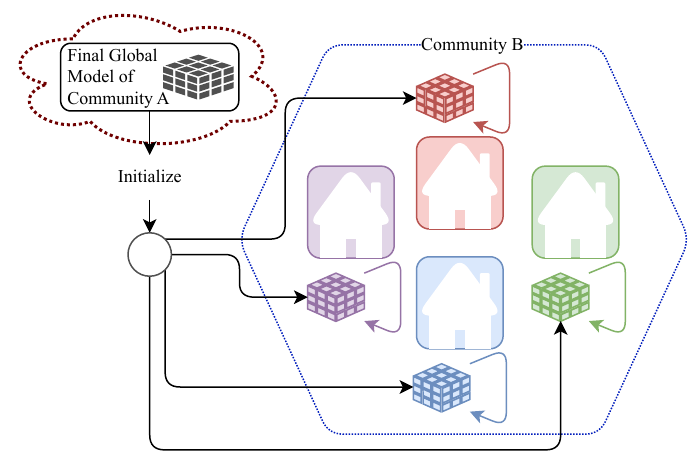}
    \caption{Diagram representing the training process by the community $b$ when choosing the FL approach.}
    \label{f:local_train_process_cmtyb}
\end{figure}

The process of training the forecast model depends on the community, and the approach taken:
\begin{description}
    \item [1. Community $a$ and $b$ - LMF:] Each building has access to its only local data, and trains locally their forecast models (there is no collaboration). Represented in Fig.~\ref{f:local_train_process};
    \item [2. Community $a$ - FLF:] Each building has access to its only local data, but using a FL approach, round by round, the server applied Eq.~\eqref{eq:FedAvg} enabling collaboration. Represented in Fig.~\ref{f:fl_implementation};
    \item [3. Community $b$ - FLF:] Similar approach to the LMF training process. However, the local models are initiated by the last global model achieved by community $a$. Represented in Fig.~\ref{f:local_train_process_cmtyb}; 
\end{description}

The FL simulations were performed using Spyder IDE 5.2.2 platform with  Python 3.9.13 64-bit in Windows 10 environment using the AMD Ryzen 9 590X 12-Core Processor with 32 GB of RAM.

\subsection{\textit{Consensus + Innovations} Approach Setup \& Initialization}  \label{s:simulations_cpi}
The setup was divided into a 5-agent system 
, as shown in Fig. \ref{f:eap_diagram}. 
The communication network bridges 4 VPPs and 1 community. 
The community has two structures as described in Section \ref{s:simulations_data}: Community $a$ with 100 buildings and Community $b$ with 25 buildings. 
Both agent-based structures are evaluated with the FL-assisted \textit{Consensus + Innovations} approach.
Three different net demand forecast methods are considered as input into the \textit{Consensus + Innovations} Approach: 
1) Real Net Demand (RND) (unknown in reality, considered as a baseline);
2) Federated Learning Forecasted (FLF) net demand;
3) Local Model Forecasted (LMF) net demand.
The net demand prediction values are obtained using the methods presented in Section \ref{s:simulations_data} and Section \ref{s:simulations_fl}.

The real-time simulation procedure for the FL-assisted \textit{Consensus + Innovations} approach is similar to the process presented in Fig. \ref{f:simulation_flow}, except for training a local model along with the FL model.
Between $t_1$ and $t_2$, the community net demand is forecasted using either the FL model or a local model, which is then integrated into the \textit{Consensus + Innovations} approach. 
To ensure a fair comparison, the hyperparameters for the \textit{Consensus + Innovations} approach are uniform across all prediction methods for the same time period, tuned specifically to the RND method. 
While convergence is achieved with all forecasting methods, there is a possibility that the agents may converge to a value that deviates from the actual demand, which will be elaborated on in Section \ref{s:results_cpi}.
This 15-minute simulation workflow is repeated for 96 intervals, simulating a full day's operation. 
For a balanced comparison, the process is simulated using the last days of April, August, and December, specifically 4/30, 8/31, and 12/31.



The \textit{Consensus + Innovations} Approach simulations were performed using PyCharm (Version 2023.2.3) platform with Python 3.6 environment on iMac (Apple M1, 2021).

\section{Results \& Discussion}  \label{s:results}

This section shows results from the simulation setup presented in Section \ref{s:simulations}. 
To this end, the forecasts of net demand from both the local and FL models are compared and discussed. 
The performance of the \textit{Consensus + Innovations} approach will be evaluated with the two forecast methods, as well as real net demand.

\subsection{Performance Metrics}  \label{s:results_metrics}

The Root Mean Squared Error (RMSE) and Mean Average Error (MAE) will be used to assess the results. To this end, $y$ will be considered as the true values and $\hat y$ as the forecasted values. Introducing $n$ as the number of data points, RMSE and MAE can be defined by \eqref{eq:metrics_rmse} and \eqref{eq:metrics_mae}, respectively. 

\begin{align}
    \label{eq:metrics_rmse}
    RMSE = \sqrt{\dfrac{\sum(y-\hat y)^2}{n}}
\end{align}

\begin{align}
    \label{eq:metrics_mae}
    MAE = \dfrac{\lvert \sum(y-\hat y) \rvert}{n}
\end{align}

The performance of the \textit{Consensus + Innovations} approach under the three net demand forecast methods is assessed using the total energy price difference. This metric studies the maximum distance between the agents' converging energy prices $\lambda_{i, \tau}^{conv}$ and the true optimal system energy price ($\lambda_{\tau}^*$), summed over a day (96 time periods), given by \eqref{eq:ep_diff}.

\begin{align}
    Energy Price Difference = \sum_{\tau=1}^{96}\left(\max_i |\lambda_{i, \tau}^{conv} - \lambda_{\tau}^*|\right), i\in I\label{eq:ep_diff}
\end{align}

The true optimal energy price is derived by solving \eqref{eq:lamStar_g2c} and \eqref{eq:lamStar_c2g} in a centralized fashion.

\subsection{Federated Learning Prediction Results}  \label{s:results_fl}

\begin{figure}[htbp]
\centering
\setlength{\abovecaptionskip}{0.cm}
\includegraphics[width=1\columnwidth]{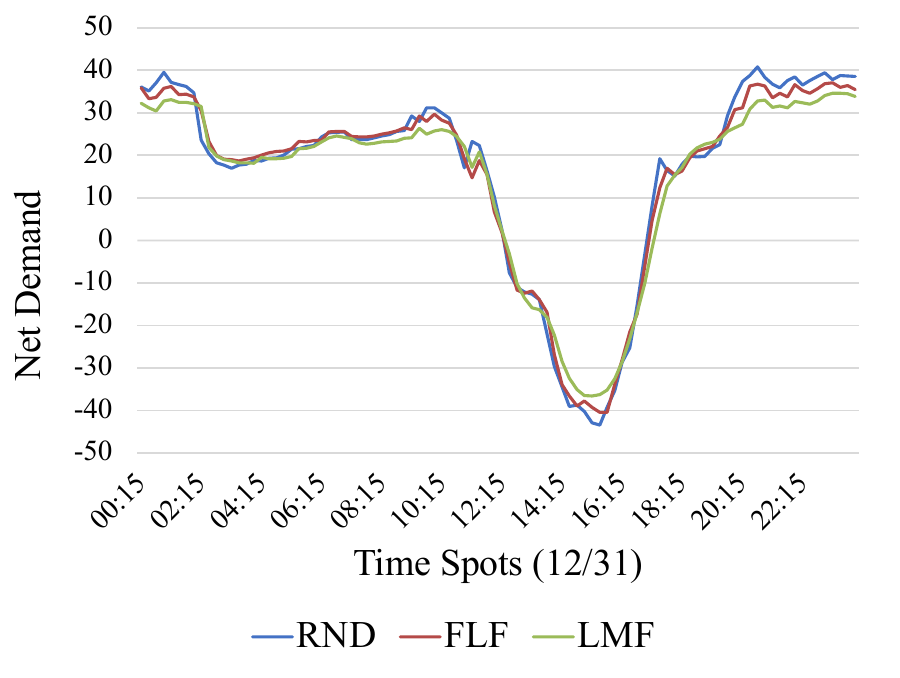}
\caption{Net demand forecast results for Community $a$ on December 31st, for RND (Real Net Demand), FLF (Federated Learning Forecasted) net demand, and LMF (Local Model Forecasted) net demand.
}
\label{f:fl_forecastResults_a}
\end{figure}

\begin{figure}[htbp]
\centering
\setlength{\abovecaptionskip}{0.cm}
\includegraphics[width=1\columnwidth]{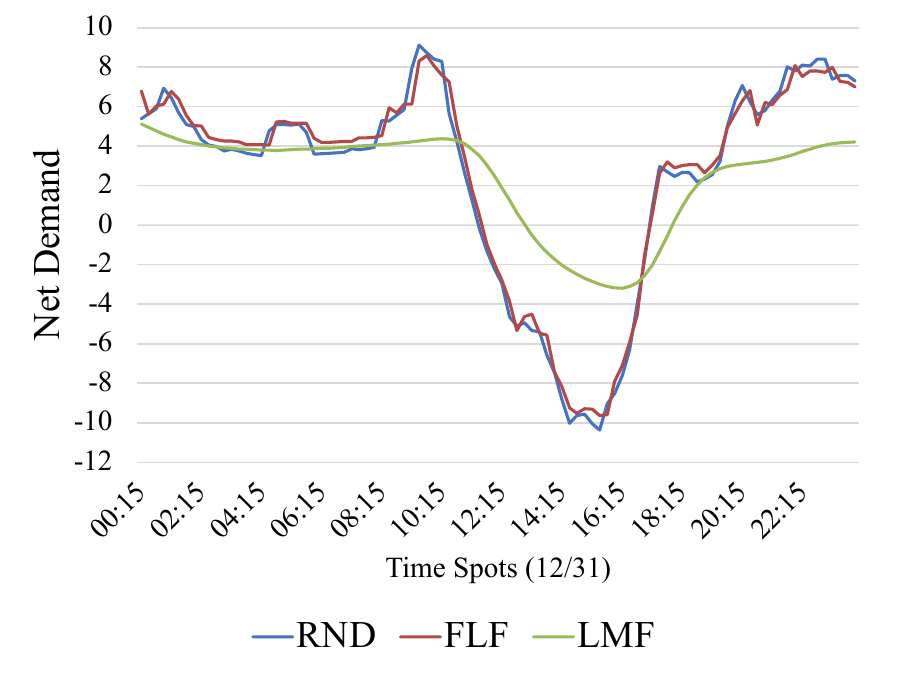}
\caption{Net demand forecast results for Community $b$ on December 31st, for RND (Real Net Demand), FLF (Federated Learning Forecasted) net demand, and LMF (Local Model Forecasted) net demand.
}
\label{f:fl_forecastResults_b}
\end{figure}

The forecast results are presented in Table~\ref{t:forecast_results}. 
In this table, \textbf{F} refers to the framework (FLF or LMF), and \textbf{M} indicates the selected month for test or train/test (from the community of 25 buildings). Additionally, the forecasted system (demand (d)/generation (g)) is shown as \textbf{S}. The number of community buildings is presented as \textbf{C}.
Based on the obtained results, FLF achieves better performance across all tests. However, as the number of test data points increases, the gap between LMF and FLF results narrows. 
Another observation is that the difference between FLF and LMF is higher in the community with 25 buildings. This happens because FLF has the advantage of using a model pre-trained by the community with 100 buildings. 
\begin{table}[htpb]
\centering
\caption{Achieved metrics for the different tests realized.}
\label{t:forecast_results}
\begin{tblr}{
  cells = {c},
  hline{1-2,14,26} = {-}{0.08em},
  hline{4,8,12,16,20,24} = {-}{dashed},
  hline{6,10,18,22} = {-}{},
}
\textbf{F} & \textbf{M} & \textbf{S} & \textbf{C} & \textbf{RMSE} & \textbf{MAE} \\
FLF & 04 & d & 100 & 0.08311 & 0.04410 \\
LMF & 04 & d & 100 & 0.13456 & 0.08410 \\
FLF & 04 & g & 100 & 0.02149 & 0.01717 \\
LMF & 04 & g & 100 & 0.08664 & 0.06712 \\
FLF & 08 & d & 100 & 0.08750 & 0.05228 \\
LMF & 08 & d & 100 & 0.09550 & 0.05837 \\
FLF & 08 & g & 100 & 0.04575 & 0.03201 \\
LMF & 08 & g & 100 & 0.08664 & 0.05934 \\
FLF & 12 & d & 100 & 0.09121 & 0.05046 \\
LMF & 12 & d & 100 & 0.10883 & 0.05957 \\
FLF & 12 & g & 100 & 0.01814 & 0.01177 \\
LMF & 12 & g & 100 & 0.03985 & 0.02912 \\
FLF & 04 & d & 25 & 0.01607 & 0.00915 \\
LMF & 04 & d & 25 & 0.03117 & 0.01966 \\
FLF & 04 & g & 25 & 0.00565 & 0.00425 \\
LMF & 04 & g & 25 & 0.03418 & 0.03110 \\
FLF & 08 & d & 25 & 0.01665 & 0.00962 \\
LMF & 08 & d & 25 & 0.15972 & 0.08524 \\
FLF & 08 & g & 25 & 0.01245 & 0.01245 \\
LMF & 08 & g & 25 & 0.02035 & 0.01634 \\
FLF & 12 & d & 25 & 0.01778 & 0.01050 \\
LMF & 12 & d & 25 & 0.03431 & 0.02718 \\
FLF & 12 & g & 25 & 0.00325 & 0.00200 \\
LMF & 12 & g & 25 & 0.02809 & 0.02158 
\end{tblr}
\end{table}
%
Figure~\ref{f:fl_forecastResults_a} and~\ref{f:fl_forecastResults_b} show the forecast results obtained by the demand system for community $a$ and $b$ in December, particularly its last day. 
It can be observed that FLF is more accurate than LMF. In other words, Figures~\ref{f:fl_forecastResults_a} and~\ref{f:fl_forecastResults_b} highlight the advantage of using a pre-trained model.


\subsection{\textit{Consensus + Innovations} Approach Simulation Results}  \label{s:results_cpi}
\begin{figure}[htbp]
\centering
\setlength{\abovecaptionskip}{0.cm}
\includegraphics[width=1\columnwidth]{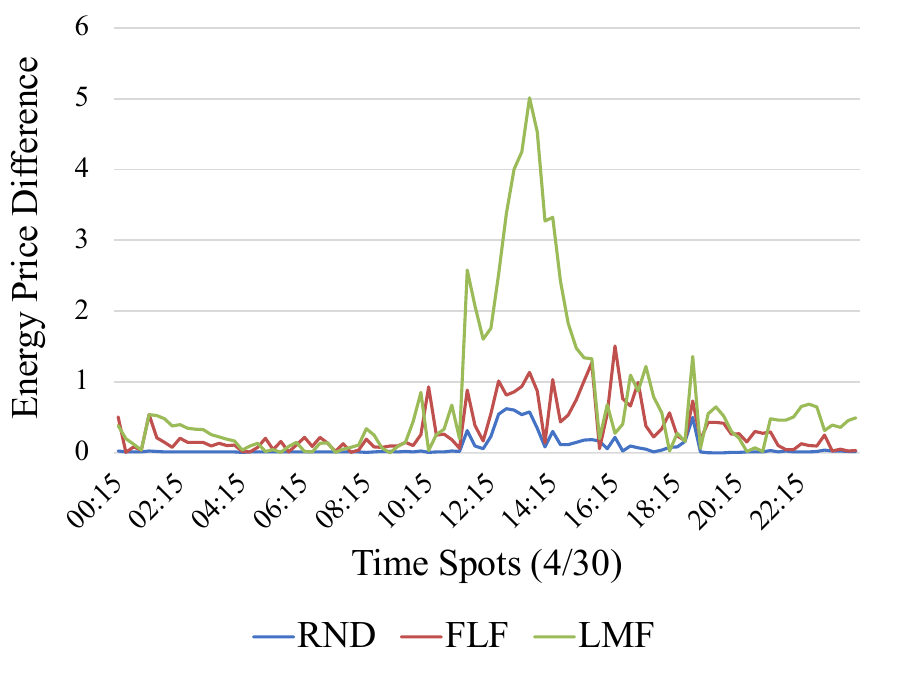}
\caption{Difference between converging energy price $\lambda$ and real optimal energy price for 96 time periods with a 15-minute interval on April 30th, after performing \textit{Consensus + Innovations} Approach for community $a$. 
}
\label{f:results_price_diff_April_100}
\end{figure}

\begin{figure}[htbp]
\centering
\setlength{\abovecaptionskip}{0.cm}
\includegraphics[width=1\columnwidth]{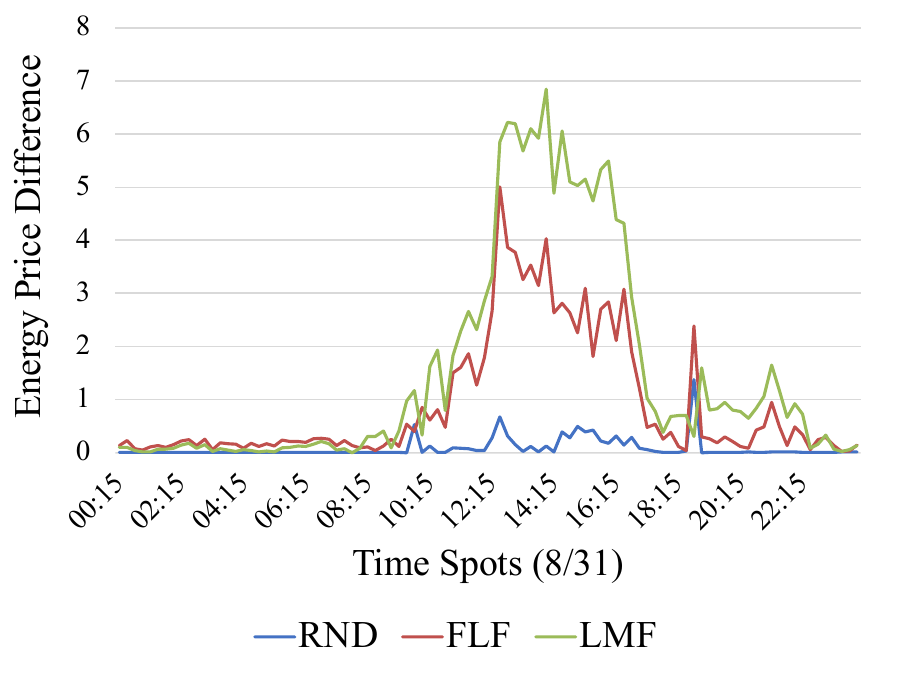}
\caption{Difference between converging energy price $\lambda$ and real optimal energy price for 96 time periods with a 15-minute interval on August 31st, after performing \textit{Consensus + Innovations} Approach for community $a$. 
}
\label{f:results_price_diff_August_100}
\end{figure}

\begin{figure}[htbp]
\centering
\setlength{\abovecaptionskip}{0.cm}
\includegraphics[width=1\columnwidth]{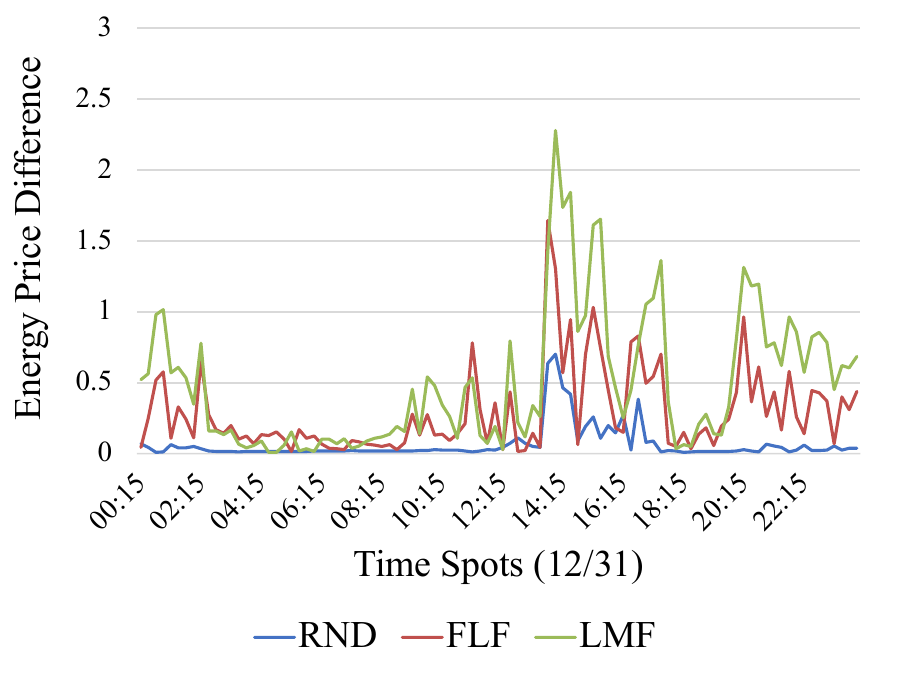}
\caption{Difference between converging energy price $\lambda$ and real optimal energy price for 96 time periods with a 15-minute interval on December 31st, after performing \textit{Consensus + Innovations} Approach for community $a$. 
}
\label{f:results_price_diff_December_100}
\end{figure}

\begin{figure}[htbp]
\centering
\setlength{\abovecaptionskip}{0.cm}
\includegraphics[width=1\columnwidth]{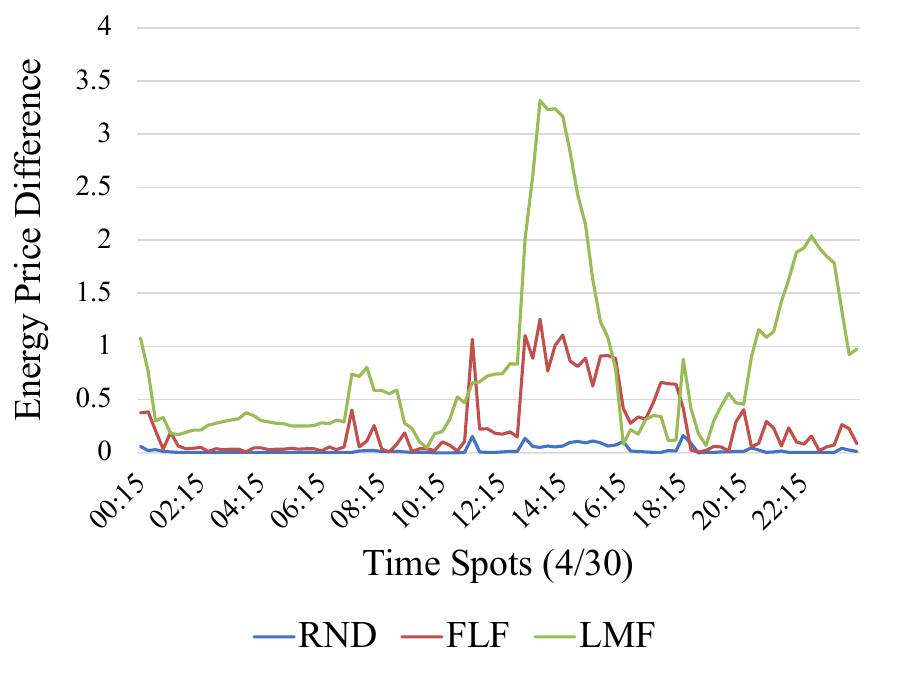}
\caption{Difference between converging energy price $\lambda$ and real optimal energy price for 96 time periods with a 15-minute interval on April 30th, after performing \textit{Consensus + Innovations} Approach for community $b$. 
}
\label{f:results_price_diff_April_25}
\end{figure}

\begin{figure}[htbp]
\centering
\setlength{\abovecaptionskip}{0.cm}
\includegraphics[width=1\columnwidth]{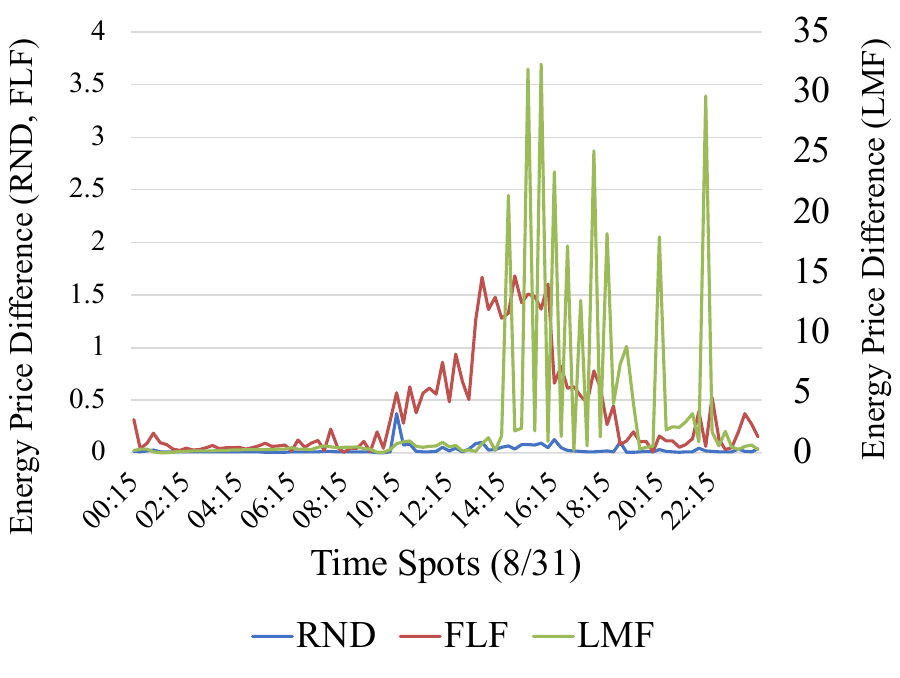}
\caption{Difference between converging energy price $\lambda$ and real optimal energy price for 96 time periods with a 15-minute interval on August 31st, after performing \textit{Consensus + Innovations} Approach for community $b$. 
The price differences for RND and FLF are shown on the left y-axis, while LMF is displayed on the right y-axis.
The maximum price difference for LMF is 32.34.
}
\label{f:results_price_diff_August_25}
\end{figure}

\begin{figure}[htbp]
\centering
\setlength{\abovecaptionskip}{0.cm}
\includegraphics[width=1\columnwidth]{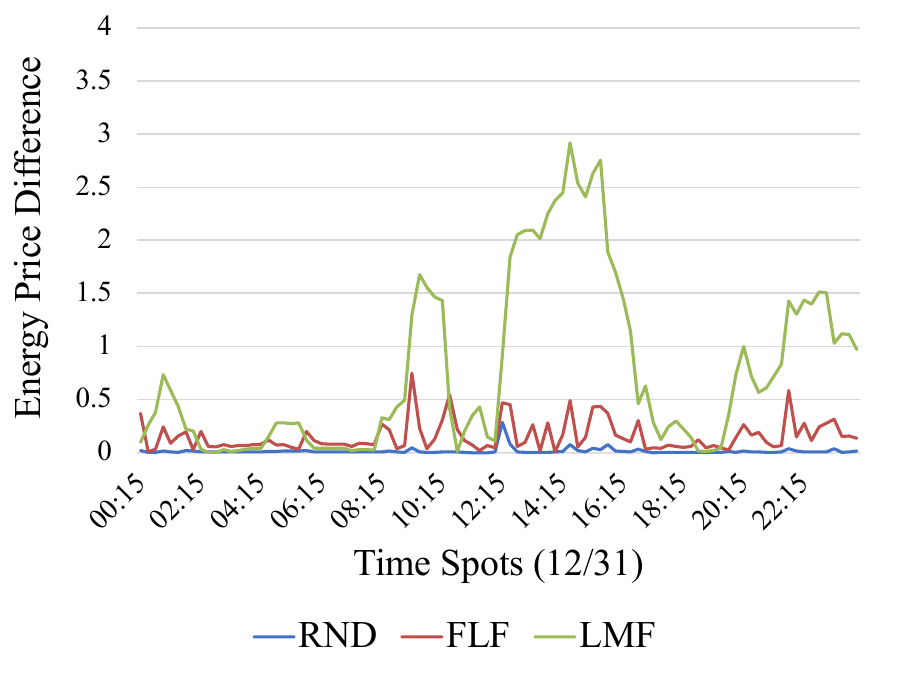}
\caption{Difference between converging energy price $\lambda$ and real optimal energy price for 96 time periods with a 15-minute interval on December 31st, after performing \textit{Consensus + Innovations} Approach for community $b$. 
}
\label{f:results_price_diff_December_25}
\end{figure}

The gap between (i) the actual optimal energy price and (ii) the agreement among local copies of energy prices (the consensus value) for two communities, and by using three different net demand forecast methods are depicted in Figures \ref{f:results_price_diff_April_100}-\ref{f:results_price_diff_December_25}. 
Additionally, the cumulative energy price difference over a day, as calculated by \eqref{eq:ep_diff}, is numerically presented in Table \ref{t:results_price_diff_summary}. 
A lower total energy price difference denotes a more precise energy price computation and superior performance. 
Notably, the Real Net Demand (RND) method demonstrates the least deviation in energy price compared to the true optimal value ($\lambda_{\tau}^*$).

\begin{table}[htbp]
\centering
\caption{Performance of the Real Net Demand (RND), Federated Learning Forecast (FLF), and Local Model Forecast (LMF). The total energy price difference is defined in \eqref{eq:ep_diff}.}
\label{t:results_price_diff_summary}
\begin{tabular*}{0.48\textwidth}{@{\extracolsep\fill}llll@{\extracolsep\fill}}
\toprule
\textbf{Testing}   & \textbf{Month}  & \textbf{Prediction}  & \textbf{Total Energy}\\
\textbf{Community} &                 & \textbf{Method}      & \textbf{Price Difference}\\
\midrule
              &           & RND  & 7.801    \\ 
              & April     & FLF  & 30.569   \\ 
              &           & LMF  & 72.127  \\
\cmidrule(r{1em}){2-4}
              &           & RND  & 8.338    \\ 
Community $a$ & August    & FLF  & 86.076   \\ 
              &           & LMF  & 143.45   \\
\cmidrule(r{1em}){2-4}
              &           & RND  & 6.074    \\ 
              & December  & FLF  & 27.606   \\ 
              &           & LMF  & 46.594   \\
\midrule
              &           & RND  & 2.455    \\ 
              & April     & FLF  & 24.542   \\ 
              &           & LMF  & 76.468   \\
\cmidrule(r{1em}){2-4}
              &           & RND  & 2.471    \\ 
Community $b$ & August    & FLF  & 36.546   \\ 
              &           & LMF  & 305.104  \\
\cmidrule(r{1em}){2-4}
              &           & RND  & 1.656    \\ 
              & December  & FLF  & 14.634    \\ 
              &           & LMF  & 73.341   \\
\bottomrule
\end{tabular*}
\begin{tablenotes}
\end{tablenotes}
\end{table}

Table \ref{t:results_price_diff_summary} shows a reduced price deviation when contrasting the FLF with the LMF, indicating its superiority. 
In specific instances, the price difference for FLF achieves up to 1.36 times greater accuracy than LMF in April for community $a$ and up to 8 times better accuracy in August for community $b$. 
Figures \ref{f:fl_forecastResults_a} and \ref{f:fl_forecastResults_a} have shown that FLF exhibits less forecast error than LMF.
This analysis confirms that net demand forecasting accuracy is a pivotal factor influencing the \textit{Consensus + Innovations} approach's performance. 
Minimizing forecast errors reduces the distributed optimization error and enhances the proposed system's reliability. 
Conversely, a high forecast error can misguide the \textit{Consensus + Innovations} updates towards erroneous values.

A comparative assessment of communities $a$ and $b$ yields additional insights. The community $a$, with its extensive data repository, effectively trains both the FL and local models. 
In contrast, community \(b\) possesses limited data samples, impacting model training. 
As Table \ref{t:results_price_diff_summary} suggests, the performance discrepancy between the FL and local models is more pronounced in the community $b$ than in $a$. 
This indicates that the FL-assisted \textit{Consensus + Innovations} approach is proficient at model transfer to similar systems with less historical data.

The results in Figure \ref{f:results_price_diff_April_100} also distinguish two error types: those from net demand forecasting and the inherent errors of the distributed iterative approach, that is the \textit{Consensus + Innovations} Approach. 
Systems with an unknown net demand (FLF and LMF) are susceptible to forecasting and distributed estimation errors. 
In contrast, known net demand scenarios (RND) predominantly face errors from the distributed iterative approach, as evidenced by the blue curve in Fig. \ref{f:results_price_diff_April_100}. 
These distributed estimation errors can be mitigated through finely-tuned hyperparameters (e.g., $\alpha$ and $\beta$ in \eqref{eq:lam_vpp_1}). 
Furthermore, the magnitude of distributed estimation error appears more noticeable in larger communities. 
Comparing the total energy price differences for RND between communities $a$ and $b$ in Table \ref{t:results_price_diff_summary}, community $a$ exhibits a higher error, attributable to its larger building count compared to community $b$.

\section{Conclusion}  \label{s:conclusion}

This paper introduces a novel FL-assisted \textit{Consensus + Innovations} framework to enhance distributed energy management in Transactive Energy Communities (TEC). 
A market mechanism was developed to optimize the utilization of surplus or deficit generation in a TEC by facilitating transactions with VPPs. 
Utilizing building data from NREL, it is demonstrated that adopting an FL model to forecast the community's net power demand significantly reduces prediction errors, even in challenging circumstances such as sparse data availability.
A \textit{Consensus + Innovations} approach was employed to enable distributed coordination among VPPs and the community.
It was observed that improving forecast accuracy decreases distributed optimization errors, thereby enhancing the reliability of the proposed framework.
The paper recognizes the compounding errors arising from the distributed optimization algorithm and the net demand forecasts. It demonstrates that integrating Federated Learning (FL) with the distributed optimization procedure can mitigate this issue.


\bmsection*{Author contributions}

\textbf{Yuhan Du}: Conceptualization, investigation, methodology, resources, visualization, writing original draft, writing review and editing;
\textbf{Nuno Mendes}: Conceptualization, investigation, methodology, resources, visualization, writing original draft, writing review and editing;
\textbf{Simin Rasouli}: Investigation, writing original draft;
\textbf{Javad Mohammadi}: Conceptualization, investigation, methodology, supervision, resources, visualization, writing review and editing;
\textbf{Pedro Moura}: Conceptualization, investigation, methodology, supervision, resources, visualization, writing review and editing.


\bmsection*{Acknowledgments}
Generative Artificial
Intelligence was used to enhance grammar and readability

\bmsection*{Financial disclosure}
This research was supported by (i) the National Science Foundation (\#2313768), (ii) the Portuguese Foundation for Science and Technology through the project ML@GridEdge (UTAP-EXPL/CA/0065/2021), (iii) the ERDF and (iv) national funds through the project EVAICharge (CENTRO-01-0247-FEDER-047196).

\bmsection*{Conflict of interest}
The authors declare no potential conflict of interest.

\bibliography{references}

\begin{thebibliography}{10}

\bibitem{TE_definition}
Council TGA. 2019. {GridWise} {Transactive} {Energy} {Framework} {Version} 1.1.
\newblock GridWise Architecture Council.
\newblock PNNL-22946 Ver. 1.1Available from: \url{https://gridwiseac.org/}.

\bibitem{du2022learning}
Du Y, Li M, Mohammadi J, Blasch E, Aved A, Ferris D, et~al.
\newblock Learning Assisted Agent-based Energy Optimization: A Reinforcement
  Learning Based Consensus+ Innovations Approach.
\newblock In: 2022 North American Power Symposium (NAPS). IEEE; 2022. p. 1--6.

\bibitem{kar2014distributed}
Kar S, Hug G, Mohammadi J, Moura JM.
\newblock Distributed state estimation and energy management in smart grids: A
  consensus + innovations approach.
\newblock IEEE Journal of selected topics in signal processing.
  2014;8(6):1022--1038.

\bibitem{kar2012distributed}
Kar S, Hug G.
\newblock Distributed robust economic dispatch in power systems: A consensus+
  innovations approach.
\newblock In: 2012 IEEE Power and Energy Society General Meeting. IEEE; 2012.
  p. 1--8.

\bibitem{li2023machine}
Li M, Mohammadi J.
\newblock Machine Learning Infused Distributed Optimization for Coordinating
  Virtual Power Plant Assets.
\newblock arXiv preprint arXiv:231017882. 2023;.

\bibitem{kargarian2016toward}
Kargarian A, Mohammadi J, Guo J, Chakrabarti S, Barati M, Hug G, et~al.
\newblock Toward distributed/decentralized DC optimal power flow implementation
  in future electric power systems.
\newblock IEEE Transactions on Smart Grid. 2016;9(4):2574--2594.

\bibitem{li2023learning}
Li M, Kolouri S, Mohammadi J.
\newblock Learning to Optimize Distributed Optimization: ADMM-based DC-OPF Case
  Study.
\newblock In: 2023 IEEE Power \& Energy Society General Meeting (PESGM). IEEE;
  2023. p. 1--5.

\bibitem{khan2022multi}
Khan MW, Wang J.
\newblock Multi-agents based optimal energy scheduling technique for electric
  vehicles aggregator in microgrids.
\newblock International Journal of Electrical Power \& Energy Systems.
  2022;134:107346.

\bibitem{nguyen2021distributed}
Nguyen TL, Tran QT, Caire R, Wang Y, Besanger Y, Luu NA.
\newblock Distributed optimal power flow and the multi-agent system for the
  realization in cyber-physical system.
\newblock Electric Power Systems Research. 2021;192:107007.

\bibitem{jasim2023consensus}
Jasim AM, Jasim BH, Mohseni S, Brent AC.
\newblock Consensus-based dispatch optimization of a microgrid considering
  meta-heuristic-based demand response scheduling and network packet loss
  characterization.
\newblock Energy and AI. 2023;11:100212.

\bibitem{mohammadi2014distributed}
Mohammadi J, Kar S, Hug G.
\newblock Distributed approach for DC optimal power flow calculations.
\newblock arXiv preprint arXiv:14104236. 2014;.

\bibitem{mohammadi2016agent}
Mohammadi J, Hug G, Kar S.
\newblock Agent-based distributed security constrained optimal power flow.
\newblock IEEE Transactions on Smart Grid. 2016;9(2):1118--1130.

\bibitem{mohammadi2016fully}
Mohammadi J, Hug G, Kar S.
\newblock A fully distributed cooperative charging approach for plug-in
  electric vehicles.
\newblock IEEE Transactions on Smart Grid. 2016;9(4):3507--3518.

\bibitem{mohammadi2023towards}
Mohammadi M, Thornburg J, Mohammadi J.
\newblock Towards an energy future with ubiquitous electric vehicles: Barriers
  and opportunities.
\newblock Energies. 2023;16(17):6379.

\bibitem{ahmadifar_development_2023}
Ahmadifar A, Ginocchi M, Golla MS, Ponci F, Monti A.
\newblock Development of an {Energy} {Management} {System} for a {Renewable}
  {Energy} {Community} and {Performance} {Analysis} via {Global} {Sensitivity}
  {Analysis}.
\newblock IEEE Access. 2023;11:4131--4154.
\newblock Conference Name: IEEE Access.

\bibitem{dinh_milp-based_2022}
Dinh HT, Kim D, Kim D.
\newblock {MILP}-based optimal day-ahead scheduling for a system-centric
  community energy management system supporting different types of homes and
  energy trading.
\newblock Scientific Reports. 2022 Oct;12(1):18305.
\newblock Number: 1 Publisher: Nature Publishing Group.
\newblock Available from:
  \url{https://www.nature.com/articles/s41598-022-22293-y}.

\bibitem{kim_convolutional_2017}
Kim P, Kim P.
\newblock Convolutional neural network.
\newblock MATLAB deep learning: with machine learning, neural networks and
  artificial intelligence. 2017;p. 121--147.
\newblock Publisher: Springer.

\bibitem{ghalehkhondabi_overview_2017}
Ghalehkhondabi I, Ardjmand E, Weckman GR, Young WA.
\newblock An overview of energy demand forecasting methods published in
  2005–2015.
\newblock Energy Systems. 2017 May;8(2):411--447.
\newblock Available from: \url{https://doi.org/10.1007/s12667-016-0203-y}.

\bibitem{chae_artificial_2016}
Chae YT, Horesh R, Hwang Y, Lee YM.
\newblock Artificial neural network model for forecasting sub-hourly
  electricity usage in commercial buildings.
\newblock Energy and Buildings. 2016 Jan;111:184--194.
\newblock Available from:
  \url{https://www.sciencedirect.com/science/article/pii/S0378778815304102}.

\bibitem{azadeh_annual_2006}
Azadeh MA, Sohrabkhani S.
\newblock Annual {Electricity} {Consumption} {Forecasting} with {Neural}
  {Network} in {High} {Energy} {Consuming} {Industrial} {Sectors} of {Iran}.
\newblock In: 2006 {IEEE} {International} {Conference} on {Industrial}
  {Technology}; 2006. p. 2166--2171.

\bibitem{sabzehgar_solar_2020}
Sabzehgar R, Amirhosseini DZ, Rasouli M.
\newblock Solar power forecast for a residential smart microgrid based on
  numerical weather predictions using artificial intelligence methods.
\newblock Journal of Building Engineering. 2020 Nov;32:101629.
\newblock Available from:
  \url{https://www.sciencedirect.com/science/article/pii/S2352710219330414}.

\bibitem{zhang_forecast_2018}
Zhang R, Feng M, Zhang W, Lu S, Wang F.
\newblock Forecast of {Solar} {Energy} {Production} - {A} {Deep} {Learning}
  {Approach}.
\newblock In: 2018 {IEEE} {International} {Conference} on {Big} {Knowledge}
  ({ICBK}); 2018. p. 73--82.

\bibitem{9960350}
Mendes N, Moura P, Mendes J, Salles R, Mohammadi J.
\newblock Federated Learning Enabled Prediction of Energy Consumption in
  Transactive Energy Communities.
\newblock In: 2022 IEEE PES Innovative Smart Grid Technologies Conference
  Europe (ISGT-Europe); 2022. p. 1--5.

\bibitem{FEKRI2022107669}
Fekri MN, Grolinger K, Mir S.
\newblock Distributed load forecasting using smart meter data: Federated
  learning with Recurrent Neural Networks.
\newblock International Journal of Electrical Power \& Energy Systems.
  2022;137:107669.
\newblock Available from:
  \url{https://www.sciencedirect.com/science/article/pii/S0142061521008991}.

\bibitem{cheng2022review}
Cheng X, Li C, Liu X.
\newblock A review of federated learning in energy systems.
\newblock 2022 IEEE/IAS Industrial and Commercial Power System Asia (I\&CPS
  Asia). 2022;p. 2089--2095.

\bibitem{9148937}
Taïk A, Cherkaoui S.
\newblock Electrical Load Forecasting Using Edge Computing and Federated
  Learning.
\newblock In: ICC 2020 - 2020 IEEE International Conference on Communications
  (ICC); 2020. p. 1--6.

\bibitem{nightingale2022effect}
Nightingale JS, Wang Y, Zobiri F, Mustafa MA.
\newblock Effect of Clustering in Federated Learning on Non-IID Electricity
  Consumption Prediction.
\newblock In: 2022 IEEE PES Innovative Smart Grid Technologies Conference
  Europe (ISGT-Europe). IEEE; 2022. p. 1--5.

\bibitem{briggs2022federated}
Briggs C, Fan Z, Andras P.
\newblock Federated learning for short-term residential load forecasting.
\newblock IEEE Open Access Journal of Power and Energy. 2022;9:573--583.

\bibitem{iot3030021}
Petrangeli E, Tonellotto N, Vallati C.
\newblock Performance Evaluation of Federated Learning for Residential Energy
  Forecasting.
\newblock IoT. 2022;3(3):381--397.
\newblock Available from: \url{https://www.mdpi.com/2624-831X/3/3/21}.

\bibitem{10.1145/3485730.3493450}
Gao J, Wang W, Liu Z, Billah MFRM, Campbell B.
\newblock Decentralized Federated Learning Framework for the Neighborhood: A
  Case Study on Residential Building Load Forecasting.
\newblock In: Proceedings of the 19th ACM Conference on Embedded Networked
  Sensor Systems. SenSys '21. New York, NY, USA: Association for Computing
  Machinery; 2021. p. 453–459.
\newblock Available from: \url{https://doi.org/10.1145/3485730.3493450}.

\bibitem{9469923}
Savi M, Olivadese F.
\newblock Short-Term Energy Consumption Forecasting at the Edge: A Federated
  Learning Approach.
\newblock IEEE Access. 2021;9:95949--95969.

\bibitem{du2023need}
Du Y, Mohammadi J.
\newblock The Need for Equitable Coordination in Multi-agent Power Systems.
\newblock In: 2023 IEEE Power \& Energy Society General Meeting (PESGM). IEEE;
  2023. p. 1--5.

\end{thebibliography}


\appendix
\bmsection*{Derivation of Analytical Solutions for The Energy Aggregation Problem}  
\label{s:appendix_ea}


To solve the energy aggregation problem \eqref{eq:g2c1}-\eqref{eq:g2c3} and \eqref{eq:c2g1}-\eqref{eq:c2g3}, the inequality constraints on generation are temporarily removed to simplify the problem \cite{kar2012distributed}. 
Starting with \eqref{eq:g2c1}-\eqref{eq:g2c3} when the community has a positive net demand, the Lagrangian function is reduced to \eqref{eq:lag1}.

\vspace{-0.4cm}
\begin{align}
    \mathcal{L} = \sum_{g=1}^{N_G}C_g^{g2c}(P_g) + \lambda \cdot \left(P_{community} - \sum_{g=1}^{N_G}P_g\right) \label{eq:lag1}
\end{align}

Deriving the first-order optimality conditions gives \eqref{eq:pLpPg} and \eqref{eq:pLpLam}.

\vspace{-0.4cm}
\begin{align}
    \frac{\partial\mathcal{L}}{\partial P_g} = 2c_{1, g}^{g2c} P_g + c_{2, g}^{g2c} - \lambda = 0 \label{eq:pLpPg} \\
    \frac{\partial\mathcal{L}}{\partial \lambda}=P_{community} - \sum_{g=1}^{N_G}P_g = 0 \label{eq:pLpLam}
\end{align}

Assume that $\lambda^*$ and $P_g^*$ give the solution to \eqref{eq:pLpPg} and \eqref{eq:pLpLam}, \eqref{eq:pg1} and \eqref{eq:lam11} can be obtained.

\vspace{-0.4cm}
\begin{align}
    P_{g}^* = \frac{\lambda^* - c_{2, g}^{g2c}}{2c_{1, g}^{g2c}} \label{eq:pg1} \\
    P_{community} - \sum_{g=1}^{N_G}\frac{\lambda^* - c_{2, g}^{g2c}}{2c_{1, g}^{g2c}} = 0 \label{eq:lam11}
\end{align}

Transform of \eqref{eq:lam11} gives the expression for $\lambda^*$,

\vspace{-0.4cm}
\begin{align}
    \lambda^* = \left(\sum_{g=1}^{N_G}\frac{1}{2c_{1, g}^{g2c}}\right)^{-1} \left(P_{community} + \sum_{g=1}^{N_G}\frac{c_{2, g}^{g2c}}{2c_{1, g}^{g2c}}\right) \label{eq:lam12}
\end{align}

The above $\lambda^*$ describes the Lagrange multiplier for agents with power generation not reaching limits (non-biding inequality constraints). If the power generation is reaching limits, the new Lagrangian function becomes \eqref{eq:lag2}.

\vspace{-0.4cm}
\begin{multline}  \label{eq:lag2}
    \mathcal{L}_2 = \sum_{g=1}^{N_G}C_g^{g2c}(P_g) + \lambda \cdot \left(P_{community} - \sum_{g=1}^{N_G}P_g\right) \\ + \sum_{g=1}^{N_G} \mu_{u, g} \cdot
    \left(P_g - \overline P_g\right) \\ + \sum_{g=1}^{N_G} \mu_{l, g} \cdot \left(- P_g + \underline P_g\right)
\end{multline}

Solving first-order optimality conditions for \eqref{eq:lag2} gives \eqref{eq:pLpPg2} and \eqref{eq:pLpLam2}.

\vspace{-0.4cm}
\begin{align}
    \frac{\partial\mathcal{L}_2}{\partial P_g} = 2c_{1, g}^{g2c} P_g + c_{2, g}^{g2c} - \lambda + \mu_{u, g} - \mu_{l, g} = 0 \label{eq:pLpPg2} \\
    \frac{\partial\mathcal{L}_2}{\partial \lambda}=P_{community} - \sum_{g=1}^{N_G}P_g = 0 \label{eq:pLpLam2}
\end{align}

The power generation for agents has three cases: 
1) within limits and not reaching both bounds ($g\notin {\Omega_{\overline{B}}\cup\Omega_{\underline{B}}}$), in this case $\mu_{u, g} = 0$ and $\mu_{l, g} = 0$ ;
2) reaching the upper bound ($g\in {\Omega_{\overline{B}}}$); 
3) reaching the lower bound ($g\in {\Omega_{\underline{B}}}$).
From \eqref{eq:pLpPg2}, the power generation becomes \eqref{eq:fooc3} - \eqref{eq:fooc5}.

\vspace{-0.4cm}
\begin{align}
    2c_{1, g}^{g2c} P_g^* + c_{2, i}^{g2c} - \lambda^* = 0, g\notin {\Omega_{\overline{B}}\cup\Omega_{\underline{B}}} \label{eq:fooc3} \\
    P_g^* = \overline P_g, g\in {\Omega_{\overline{B}}} \label{eq:fooc4} \\
    P_g^* = \underline P_g, g\in {\Omega_{\underline{B}}} \label{eq:fooc5}
\end{align}

Substituting $P_g$ in \eqref{eq:pLpLam2} by \eqref{eq:fooc3}-\eqref{eq:fooc5} provides \eqref{eq:lam21}.

\vspace{-0.4cm}
\begin{align}
    P_{community} - \sum_{g\notin {\Omega_{\overline{B}}\cup\Omega_{\underline{B}}}}\frac{\lambda^* - c_{2, g}^{g2c}}{2c_{1, g}^{g2c}} - \sum_{g\in {\Omega_{\overline{B}}}}\overline P_g - \sum_{g\in {\Omega_{\underline{B}}}}\underline P_g = 0\label{eq:lam21}
\end{align}

Thus, the generalized form of (\ref{eq:lam12}) becomes \eqref{eq:lam22}.

\vspace{-0.4cm}
\begin{multline} \label{eq:lam22}
    \lambda^* = \left(\sum_{g\notin {\Omega_{\overline{B}}\cup\Omega_{\underline{B}}}}\frac{1}{2c_{1, g}^{g2c}}\right)^{-1} \left[P_{community} - \sum_{g\in {\Omega_{\overline{B}}}} \overline P_g \right. \\
    \left. \quad - \sum_{g\in {\Omega_{\underline{B}}}} \underline P_g + \sum_{g\notin{\Omega_{\overline{B}}\cup\Omega_{\underline{B}}}}\frac{c_{2, g}^{g2c}}{2c_{1, g}^{g2c}}\right]
\end{multline}

Thus, we can find analytical solution for $\lambda^*$ and $P_{G_i}^*$ in the centralized energy aggregation problem \eqref{eq:g2c1}-\eqref{eq:g2c3} by \eqref{eq:fooc3}-\eqref{eq:fooc5} and \eqref{eq:lam22}. 
The derivation process for \eqref{eq:c2g1}-\eqref{eq:c2g3} when the community has a negative net demand is similar, with analytical solution \eqref{eq:fooc4}, \eqref{eq:fooc5}, \eqref{eq:fooc6}, and \eqref{eq:lam3}. 

\vspace{-0.4cm}
\begin{align}
    2c_{1, g}^{c2g} P_g^* + c_{2, i}^{c2g} - \lambda^* = 0, g\notin {\Omega_{\overline{B}}\cup\Omega_{\underline{B}}} \label{eq:fooc6}
\end{align}

\vspace{-0.4cm}
\begin{multline} \label{eq:lam3}
    \lambda^* = \left(\sum_{g\notin {\Omega_{\overline{B}}\cup\Omega_{\underline{B}}}}\frac{1}{2c_{1, g}^{c2g}}\right)^{-1} \left[- P_{community} - \sum_{g\in {\Omega_{\overline{B}}}} \overline P_g \right. \\
    \left. \quad - \sum_{g\in {\Omega_{\underline{B}}}} \underline P_g + \sum_{g\notin{\Omega_{\overline{B}}\cup\Omega_{\underline{B}}}}\frac{c_{2, g}^{c2g}}{2c_{1, g}^{c2g}}\right]
\end{multline}





\end{document}